\documentclass[pra,12pt,tightenlines,nofootinbib,floatfix]{revtex4} 
\pdfoutput=1
\usepackage{amsmath}
\usepackage{graphics, graphicx}
\usepackage{epsfig}
\usepackage{ulem}
\usepackage{color}
\usepackage{datetime} 

\begin{document} 

\title{The Genesis of the No-Boundary Wave Function of the Universe}

\author{James  Hartle}
\affiliation{Department of Physics, University of California, Santa Barbara, California, 93106, USA} {\affiliation{ Santa Fe Institute, \\ 1399 Hyde Park Road,  Santa Fe, New Mexico  87501, USA.}  
 \bibliographystyle{unsrt}



\begin{abstract}
{Brief recollections by the author about how he and Stephen Hawking arrived at the theory of the No-Boundary Quantum State of the Universe.}
\end{abstract}




\maketitle


\section{Introduction}
\label{intro}
It is an inescapable inference from the physics of this century and the last one
that we live in a quantum Universe, a Universe in which the process of prediction is carried out using that predictive 
framework we call quantum mechanics on all scales, from those of the elementary particles to those of the Universe
itself. 

A quantum Universe must have a quantum state. 
A theory of that state is a necessary part of any `final theory' together with a fundamental theory of the quantum dynamics of spacetime geometry and matter fields.
The no-boundary quantum state of the Universe (NBWF) put forward by Stephen Hawking and myself in 1983 is a leading candidate for a theory of this state \cite{NBWF,WINBWF}.

By now there are other candidate theories for the Universe's quantum state e.g. \cite{Vilk}. This paper is not a review of those other theories. It's not even a review of the NBWF! Rather the paper contains only my brief recollections of how Stephen and I got to the NBWF together with a few references relevant to those recollections.  For that reason there are no equations in this paper. The reader can find plenty of them in the cited papers.

\subsection*{Caveats} 
\label{caveats}  

There are obvious caveats: Stephen is no longer with us.  I am therefore relying on my memory of events nearly forty years ago with all the risks that implies. I have a few records of our calculations but they do not explain motivation. The story of the genesis of the NBWF is written in the papers Stephen and I wrote and the recollections of our interactions  that I have.

This is not a review of the NBWF's successes, nor of the significant impact it had on our understanding of our Universe, e.g,  \cite{HarClassU}, nor the reformulation  and generalization of textbook quantum mechanics that it motivated e.g. \cite{Gel-HarLight}
Rather it is a personal recollection of how Stephen and I came to propose the NBWF as a theory of our Universe's quantum state.

One caveat that I don't have, concerns any exaggeration of the role that I played in this story. For most of the period after the NBWF came out I made it a point to minimize my contributions in order not to weaken any case there might be for a Nobel Prize for Stephen. For example, I made it a point to refer to the NBWF as ``Hawking's wave function of the Universe''. If I did that when Stephen was in the audience he would correct me by by mentioning my name --- typically loudly. Now I just use the name for the NBWF that Stephen later proposed--- the no-boundary wave function of the Universe.

\section{Working in  Cambridge, UK}
\label{cambridge} 

My association with Stephen began began nearly fifty years ago during a long visit to Fred Hoyle's Institute of Theoretical Astronomy (as it was known then). In residence were people like Brandon Carter, Martin Rees, Paul Davies, and Stephen Hawking --- colleagues with whom I have maintained lifelong personal and scientific contacts.  

I was a professor of physics at the University of Chicago 1981-83. Chicago required only two quarters of teaching out of a three quarter year. I used the extra quarter to visit the Department of Applied Mathematics and Theoretical Physics in Cambridge (DAMTP) where Stephen had his office. (I claimed I was visiting for  better weather. And indeed the weather in Cambridge UK was typically better than the weather in Chicago at the times of those visits.) Summers were another time to visit. There were many lively discussions during  those visits, not only with Stephen, but also with the other scientists there and at the Institute of Theoretical Astronomy. 

I remember vividly the meeting with Stephen in Santa Barbara in the summer of '82. It was there, at the.Institute for Theoretical Physics, that  Stephen and I put  two important pieces of the Wave Function of the Universe together. One piece was my work on defining quantum states in cosmology by path integrals \cite{HarQeffEU}. The other was Stephen's idea of a compact Euclidean beginning \cite{Pontifical}. Our joint paper `The Wave Function of the Universe' came out in July '83 \cite{NBWF}.   

\section{Stephen and Cosmology} 
\label{stephen-cosmo}
Stephen's deep interest in cosmology ran from his first published paper ``Occurrence of Singularities in Open Universes'' in 1965 to his last paper with Thomas Hertog ``A Smooth Exit from Eternal Inflation'' in 2018 \cite{HawkHer}. There were many more in between \cite{HawkRoySoc}. 
Stephen sought to understand the whole Universe in scientific terms. As he said famously ``My goal is simple. It is a complete understanding of the Universe.''
Pursuing this goal led to the NBWF.  

The singularity theorems proved by Stephen, Roger Penrose, and others showed conclusively that the classical Einstein equation implies that the Universe began in a big bang e.g.  \cite{Clarke-sing} . But the singularity theorems also showed that this beginning could not be described by a classical spacetime geometry obeying the Einstein equation with three space and one time direction at each point. Rather the singularity  theorems  showed something more sweeping: The classical Einstein equation breaks down at the big bang. The classical extrapolation into the past showed that near the big bang, energy scales would have been reached at which spacetime geometry would fluctuate quantum mechanically, non-classically, and be without definite value.  Quantum spacetime would be central to understand our Universe.  

Earlier joint work of Stephen and me   demonstrated the power of Euclidean geometry to help understand the quantum Hawking radiation from evaporating black holes [1]. It was therefore natural to try to use similar techniques to describe the quantum birth of the universe. Stephen first put forward a proposal along these lines at a conference in the Vatican in 1981, where he suggested that the universe began with a regular Euclidean geometry having four space dimensions that made a quantum transition to a Lorentzian geometry with three space and one time dimension that we have today [2].


\section{Working in Quantum Cosmology}
\label{jimQC} 
The late '70s and early '80s of the last century saw a lot of work on the production of elementary particles in time varying background spacetimes, especially those describing a classical expanding universe. The prediction of radiation from black holes  by Stephen was a surprising and particularly inspiring example. See e.g. \cite{Wald} for many details. 

I reasoned that if particles were created by the expansion of the Universe their back-reaction on the geometry might affect how the Universe evolves in significant ways. I wasn't  alone in this thought.  I addressed the question in a number of calculations  e.g. 
\cite{HarQeffEU,HarPIQpartprod,HarHawk2}. 



The important point for the genesis of the NBWF is not the results of these calculations of mine. What was important is that I was working on the implications of quantum mechanics for realistic cosmology in a framework that effectively involved both quantum dynamics and quantum states. I was ready for the NBWF. 

A joint paper with Stephen that influenced the genesis of the NBWF concerned the derivation of the Hawking temperature by path integral methods \cite{HarHawk2}. I think this was the first paper of Stephen's to use path integrals in cosmology and was possibly his first introduction to them. Further, the paper calculated the result by going into the complex time plane --- i.e it employed Euclidean geometry.

\section{A quantum state of the Universe defined by a Euclidean path integral.}
\label{eucfnal} 
If the Universe is a quantum mechanical system it must have a quantum state. This state of the Universe is not something discoverable by work in the lab. The quantum state must be part of the fundamental theoretical structure --- a new law of nature. 
 

I had three reasons for thinking that the quantum state of the Universe might be given by a Euclidean path integral. 
 
The first reason was that path integrals were a standard method of defining quantum states in non-relativistic quantum mechanics, as in the two-slit experiment, and, in more detail, in most quantum mechanics texts, as well as in earlier work by Stephen and me \cite{HarHawk2}. 
 
The second reason was that observations showed that the Universe was simpler earlier than it is now --- more homogeneous, more isotropic, and more nearly in thermal equilibrium. Such simplicity was characteristic of the ground states of many familiar non-relativistic systems. For many familiar non-relativistic systems, ground states could be calculated not only as a state of lowest energy but also as a Euclidean path integral. The constraints of general relativity show that there is no state of lowest energy for the closed Universes that I had assumed, simply because there was no notion of total energy for those cases. But the Universe's quantum state could still be defined by a Euclidean path integral. 
 
The third reason was that earlier joint work of Stephen and me had demonstrated the power of Euclidean geometry to help understand the quantum Hawking radiation from evaporating black holes \cite{HarHawk2}. It was therefore natural to try to use similar techniques to describe the quantum birth of the universe. Stephen first put forward a proposal along these lines at a conference in the Vatican in 1981, where he suggested that our Universe began with a regular Euclidean geometry having four space dimensions that made a quantum transition to a Lorentzian geometry with three space and one time dimension that we have today \cite{Pontifical}. 
 
Stephen's reason for a Euclidean integral  I think arose from his long standing interest in how the Universe began. His singularity theorems showed that it couldn't begin with a Lorentzian geometry. Therefore, he said, in a later talk ``Jim and I thought  that therefore it must begin with a Euclidean one'' --- classic Hawking.
 
 



 \section{The NBWF Comes Out}
 \label{comingout}  
 
In 1983 Stephen Hawking and I published our proposal for the no-boundary wave function of the Universe \cite{NBWF,WINBWF} defined in terms of a path integral over Euclidean histories of cosmological geometry and matter fields. At one stroke we were working in quantum cosmology. To support the idea of a state of the Universe defined by a Euclidean path integral, I showed how the analogous construction in gravity linearized about flat space led to the correct ground state \cite{HarLin}. Later work with Jonathan Halliwell \cite{Hal-Har90,Hal-Har91} discussed the contour of integration in the defining functional integral and showed how the result satisfied the operator constraints of quantum general relativity.

Early predictions of the NBWF for cosmology were obtained using semiclassical algorithms posited in analogy with the WKB approximation in non-relativistic quantum theory. They were successful in explaining homogeneity and isotropy of the Universe observed on large distance scales and the spectrum of its early density fluctuations that lead to today's observations of the cosmic microwave background radiation and the large scale distribution of the galaxies. The NBWF worked! 
Later work would suggest a new quantum framework for making cosmological predictions called Decoherent or Consistent Histories quantum mechanics \cite{Gel-HarLight}.

\section{What's In a Name} 
\label{name} 
The quantum state defined by the wave function of the Universe introduced in the paper \cite{NBWF} by Stephen and me is now usually referred to as the {\it No-Boundary Quantum State of the Universe.} or sometimes just as the Hartle-Hawking State (my blushes).  Either works. The No-Boundary name seems natural because the saddle points of the semiclassical approximation to the defining integral are compact with only one boundary and {\it no other boundary}, e.g Figure 5. However the original paper paper \cite{NBWF} refers only to the `wave function of the Universe'. 

Stephen later proposed the better name `The No-Boundary Quantum State' while he, Thomas Hertog, and I were working to extract its testable predictions, e.g. \cite{HarClassU} and a fair number more. The author is sticking to Stephen's choice --- `The no-boundary quantum state of the Universe'.

\
\section {Ideal Steps to the NBWF}
\label{steeps} 
This section contains a hypothetical list of steps that, if followed, would lead from classical cosmology to the NBWF. This is {\bf not a historical list of steps taken} by Stephen and me.  Our route covered the same ground but in a less systematic way characteristic of making a discovery. I display the list here rather as a way to illustrate and highllght our accomplishments however we made them.

\begin{itemize}

\item{1}. Understanding that the singularity theorems of Penrose, Stephen, and others imply that our Universe started in a big-bang where the laws of classical spacetime (e.g the Einstein equation) break down  e.g. \cite{Clarke-sing}.

\item{2}.  Understanding that step (1) implies that the beginning of the Universe is quantum mechanical in an essential way and not merely a classical boundary condition.

\item{3}. Understanding that if the Universe is a quantum mechanical system it has a quantum state --- a wave function of the Universe (WAVU). That state effectively contains the information about how the Universe began. 

\item{4} Understanding what results of our current observations a WAVU predicts with significant probability. What does it predict with significant probability will happen in the future and what did happen in the past? In short, what does the state predict for the history of our Universe from the distant past to the far future.
\end{itemize} 

Stephen and I could be said to have understood these questions and provided answers to these with the NBWF.  But that is not the case. It was just as much a challenge to find the right questions to ask, as to find the answers.

\section{Early Reaction} 
\label{earreact} 
We were not the first scientists to discuss quantum cosmology, e.g. \cite{misner,halliwell-bib}. But my impression is that at the time the idea of a wave function of the Universe was a novelty both for scientists and for lay-persons interested in science. After all, a quantum state was usually something created in a laboratory. But the NBWF is not something produced by us. It exists as a candidate for a fundamental physical law to be discovered by us. However, I did not find it that difficult to explain it to both categories of interested parties.        

The NBWF was rapidly taken up by a number of scientists notably by Stephen's students, 
Jonathan Halliwell for example.\cite{HalHawkstr} 

\section{Classical Spacetime} 
Stephen and I first developed the NBWF using a conformally invariant scalar field for the matter degrees of freedom and that is what appears  in the first paper \cite{NBWF}. It wasn't completely clear from the result that classical cosmological spacetime was predicted. However we subsequently found, assuming a massive scalar field, that the NBWF did correctly predict classical cosmological spacetime \cite{Hawk84}. 
Only Stephen's name appears on that paper. That came about this way:  As mentioned above in this period I oscillated between Cambridge and the University of Chicago where I taught physics. Returning to Cambridge to work on the NBWF I found that Stephen had already completed  a minisuperspace model with a massive scalar field that clearly predicted classical cosmological spacetime. As a team he wanted the paper to be joint and suggested that my name appear as well.  But in those days I had the idea that I should not put my name on a paper if I hadn't done a significant fraction of the calculations. I insisted. Despite Stephen's urging it appeared under Stephen's name alone.

\section{afterwards -- Physics}
\label{afterwards}

\subsection{Further Development} 

Drawing on the background discussed above the idea of the NBWF was further developed over a period of time by Stephen's students and postdocs and my students and postdocs in Santa Barbara.  It was also developed in the many places where Stephen and I happened to be together  --- Cambridge, Cal. Tech. Boston, Chicago, Santa Barbara, etc.

\subsection{Fundamental Theories}

The NBWF has had a major impact on our efforts to understand our quantum Universe. In particular, it showed explicitly that what is called a final (or fundamental) theory, sometimes called `a theory of everything', has two parts: First, a theory of the Universe's quantum dynamics specifying regularities in time.  And second, a theory of the Universe's quantum state specifying mostly regularities in space. The latter may be unified with the dynamical theory, as the NBWF is. Further, the NBWF showed explicitly how to extract testable predictions for the results of our observations on the largest scales of space and time.
 
 \subsection{Cosmological Observations}
 \label{cosmo-obs} 
The NBWF is  a major part of our understanding of features of our quantum Universe. A short list of such features would include: the Universe's limited regime of approximately classical spacetime. \cite{HarClassU}; its approximate homogeneity and isotropy on scales above several hundred megaparsecs \cite{HalHawkstr}; the emergence of Copenhagen quantum theory in measurement situations as an approximation later in the Universe \cite{HarDHCH}.
   
   \subsection{Arrows of Time}
   \label{arrows}  
    The visible universe exhibits several time asymmetries called arrows of time. There is the fluctuation arrow defined by the increase in deviations from homogeneity and the formation of structures such as nucleated bubbles, galaxies, stars, planets, and biota. There is the arrow defined by the retardation of electromagnetic radiation. There is the psychological arrow defined by our distinction between past, present, and future. There is the thermodynamic arrow defined by the tendency of presently isolated systems to evolve toward equilibrium in the same direction of time, and the general tendency to increase of an entropy defined by a coarse graining related to conserved quantities. Stephen had a keen interest in the arrows of time going back to his Ph.D. thesis. The NBWF provides an understanding of them.
    
Today's candidates for a fundamental theory of quantum dynamics are generally time neutral --- not preferring one direction in time over any other. The Universes's arrows of time emerge from time-asymmetric boundary conditions on the time-neutral dynamical theory.
In particular,  the arrows of rime described above can be seen to emerge from an initial no-boundary stare and a final state of ignorance (e.g one represented by a unit density matrix.)  \cite{Hawketalarrows,HarQCCG,HarHerarrows}. In this way the NBWF leads to an understanding of arrows of time.
  
 \subsection{Simplicity and Complexity}  
\label{simpcomplex} 

The evidence from observations is that our early Universe a little after the big-bang was remarkably simple (in an intuitive sense) --- approximately homogeneous, isotropic and in near thermal equilibrium. This early simplicity is implied by the NBWF. Very schematically,  as made explicit in \cite{NBWF}, it turns out that the NBWF probabilities for early fluctuations away from symmetry and simplicity are small \cite{HalHawkstr}. 

Very roughly, fluctuations away from symmetry cost Euclidean action $I$ and the NBWF damps these as $ \exp[{-(I / \hbar}]$. But the NBWF  also leaves the smaller fluctuations away from simplicity that grew under the action of gravitational attraction to make the more complex Universe we see today \cite{HalHawkstr}. 
 
 \subsection{Quantum Mechanics}
 \label{impact-qm}  The standard textbook `Copenhagen' formulation of the quantum mechanics of measurement situations must be generalized in order to to calculate the realistic predictions of the NBWF. Those generalizations occupied the author for more than a decade after the NBWF. I cite only two representative papers \cite{HarGQMspacetime}, \cite{Gel-HarLight}.
 
The NBWF was the start  of a significant intellectual transition from an understanding of quantum theory as a theory in a given, fixed, classical background spacetime of the Universe to understanding it as the probabilistic origin of the classical spacetime itself. 
 
\subsection{The Impact on Me}
\label{impactme} 
The significant impact of the NBWF on my scientific work can be judged by the small number of my papers that {\bf do not} mention the NBWF. 

Further impact of the NBWF came in the form of  interaction, expositions, and collaboration with distinguished scientists, and students of many persuasions. Thomas Hertog is certainly the most notable of these. There was wide interest. Perhaps this was because the NBWF opened up a new area of physics. But perhaps also because the NBWF enabled colleagues to participate in fundamental physics in an intuitively accessible way in answering some of the its greatest questions. 

All this allowed me to work on questions in fundamental physics as I always aspired to do without having to become a foot-soldier in some subject's army of investigators. 

Even now, late in life, I take great satisfaction in thinking of the genesis of the NBWF and how, because of that, I was able to work on something like an equal basis with one of the great figures of the age. 

But perhaps my greatest satisfaction came from the things Stephen said about the NBWF.  During a joint visit to France, Stephen told me  that the NBWF was the ``best thing either of us had done.'' In other places he said that the NBWF was his greatest achievement. 





\subsection{Memorablia} 
\label{memorabilia}

In this final section I include a few items that are not necessary to understand the argument but may help with it or amuse a reader.


\subsection{Hawking Papers with me} \begin{itemize} 

\item {}  Solutions of the Einstein-Maxwell Equations with Many Black
Holes (with S. W. Hawking), Comm. in Math Phys., 26, 87-101, 1972.

\item {}  Energy and Angular Momentum Flow into a Black Hole (with S.
W. Hawking),  Comm. Math. Phys., 27, 283-290, 1972.

\item {}  Path Integral Derivation of Black Hole Radiance (with S. W.
Hawking), \prd 13, 2188-2203, 1976.

\item {}  Wave Function of the Universe (with S. W. Hawking), \prd {\bf 28},
2960-2975, 1983. 

\item{} The No-Boundary Measure of the Universe (w/ S.W. Hawking and T. Hertog),
{\sl Phys. Rev. Lett.},  {\bf 100}, 202301 (2008), arXiv:0711:4630.

\item{} Classical Universes of the No-Boundary Quantum State, (w. S.W. Hawking and Thomas Hertog), {\sl Phys. Rev. D} {\bf  77}, 123537 (2008), arXiv:0803:1663.

\item{}  The No-Boundary Measure in the Realm of Eternal Inflation (w. S.W. Hawking and T. Hertog), {\sl Phys. Rev. D}, {\bf 82}, 063510 (2010);  arXiv:1001:0262.


\item{}  Vector Fields in Holographic Cosmology, (with S.W.~Hawking and T.~Hertog), {\sl JHEP11}  (2013) 201,  arXiv:1305.719, http://dx.doi.org/10.1007/JHEP11(2013)201. 


\item{} Accelerated Expansion from Negative $\Lambda$, (with S.W.~Hawking, and T.~Hertog), arXiv:1205.3807.  
\end{itemize} 

\subsection*{}

\begin{figure}[t]
\includegraphics[width=7in]{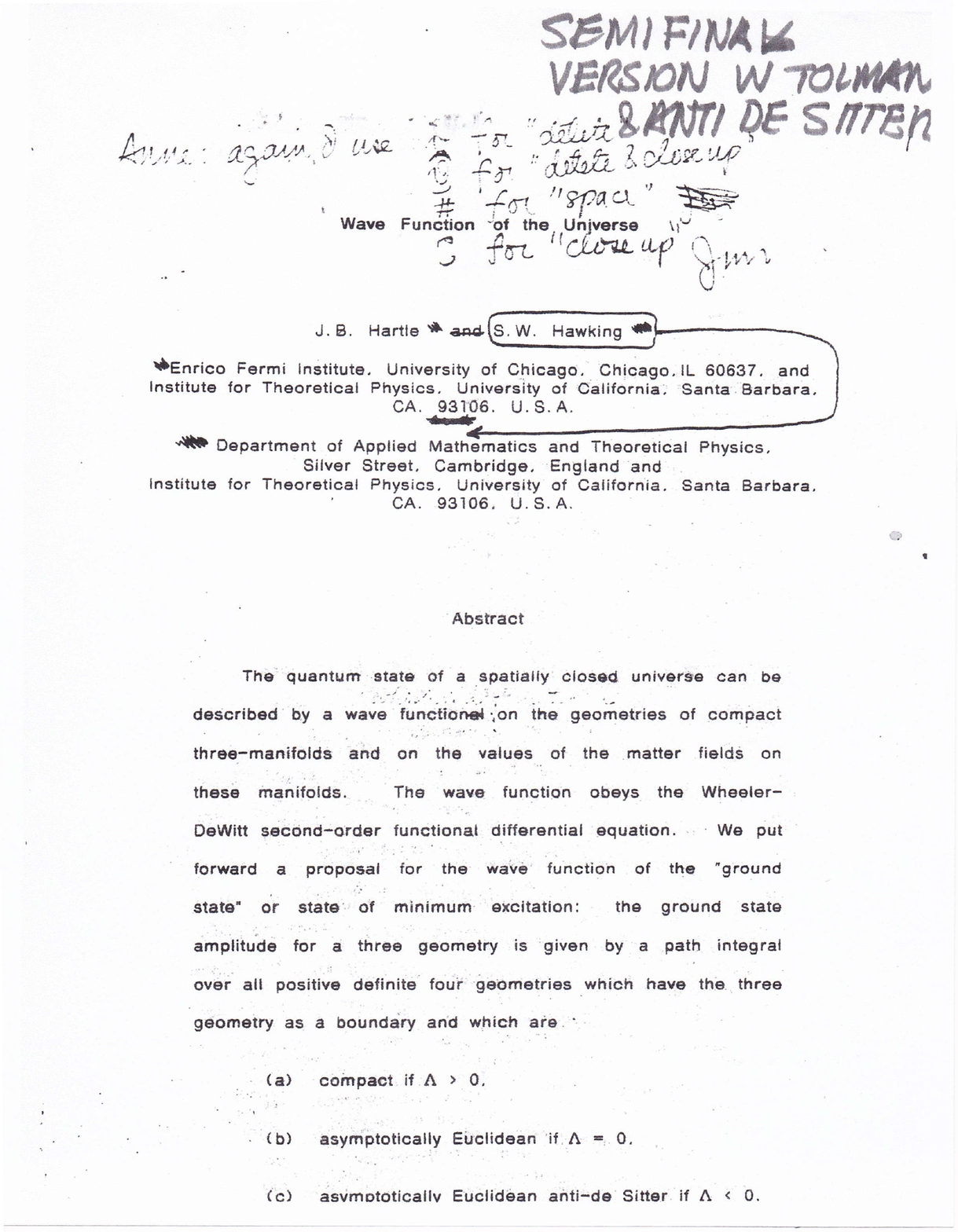}  
\caption{The first page of the typed ms of the NBWF paper for sending to the Physical Reviw unfortunately marked up. }
\label{draftp1}
\end{figure}

\begin{figure}[t]
\includegraphics[width=8in]{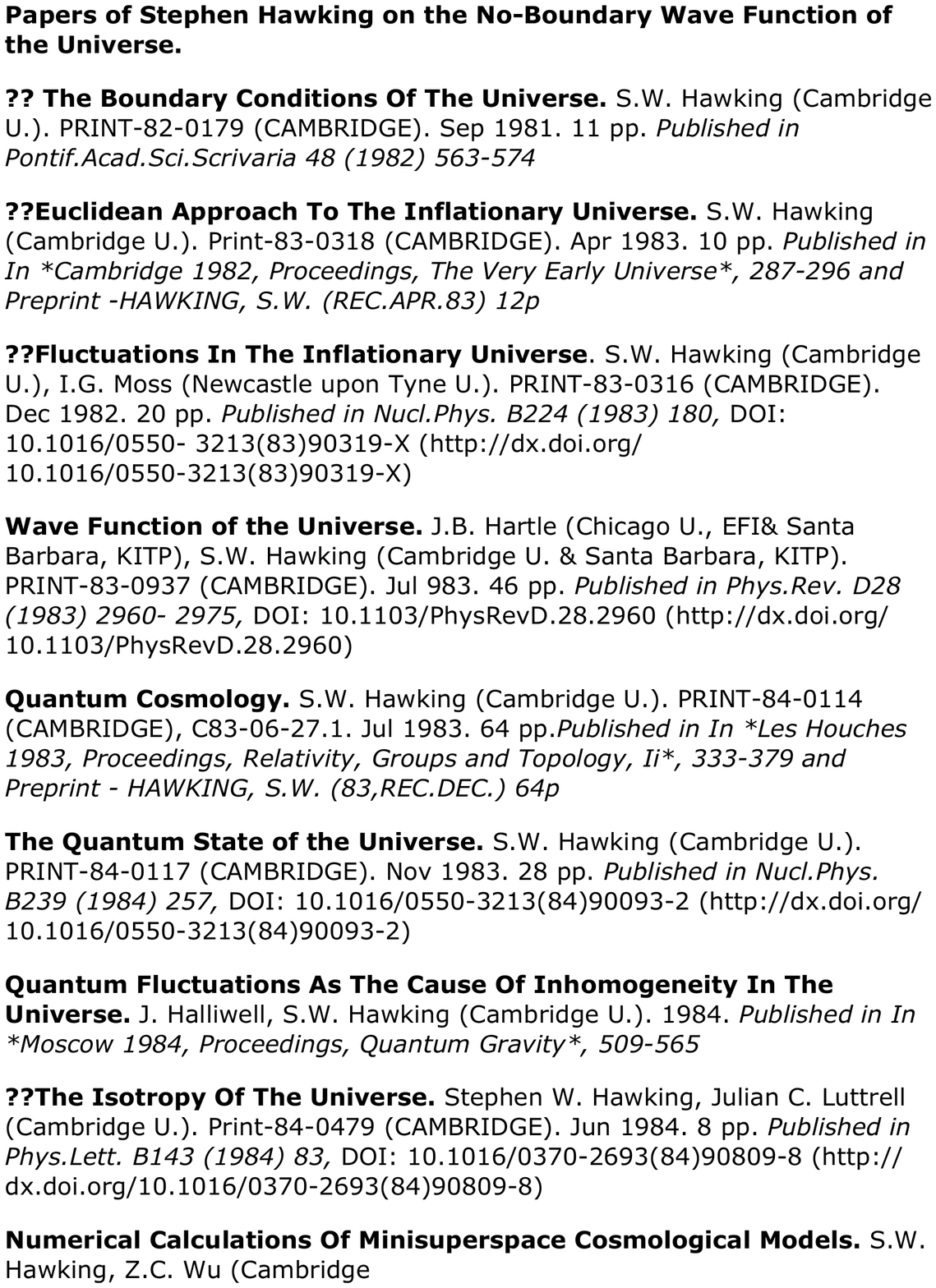}  
\caption{This is a list of papers by Hawking and  collaborators at approximately the time of genesis of THE NBWF }
\label{Hawk-NBWF  Papers}
\end{figure}


\begin{figure}[h]
\includegraphics[width=5in]{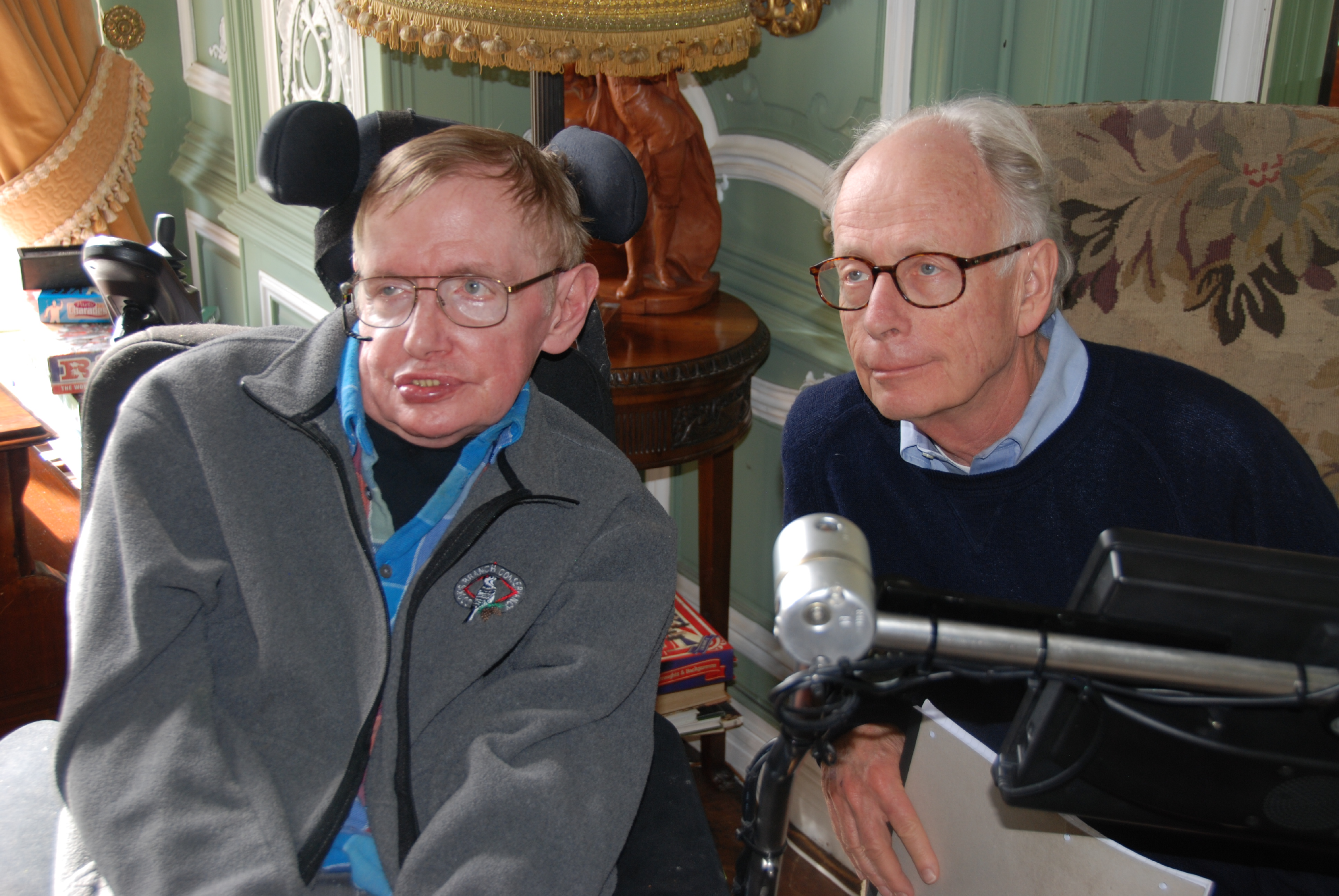}  
\caption{ Hawking  and Hartle working together in 2014}
\end{figure} 


\subsection{Working with Stephen} 

I've often been asked how I and other colleagues were able to work with Stephen. I've written elsewhere about what it was like to work with him \cite{HarWWS} , 
But Picture 4 on p13 from when Stephen, Thomas Hertog and I  were working together at George Mitchell's Cook's Branch Conservatory near Houston, Texas, gives some idea of what was involved. Working with Stephen was not significantly different from other collaborations in the usual give and take. Perhaps it was a little slower but then there were fewer mistakes and we benefited from Stephen's profound intuition.

We discussed our ideas together. If equations were needed we wrote them on the blackboard. If Stephen wanted to correct an equation or offer a new one he would describe it verbally and Thomas or I would write it on a blackboard. 

\subsection{Conclusion: A great problem} 
\label{greatprob}
It's been said that the signature of a great problem  is one that  leads to further great problems. It would I think be difficult to find a clearer example of this  than the search for the quantum state of the Universe and the understanding of the no-boundary wave-function of the Universe. 
Indeed it would not be an exaggeration to say that I have spent a large part of my modest career working on the NBWF and the great problems it has led to, confident  that it will lead to many more because of the vast sweep and scope of phenomena that it can be applied to in a manageable way.

\begin{figure}[h]
\includegraphics[width=5in]{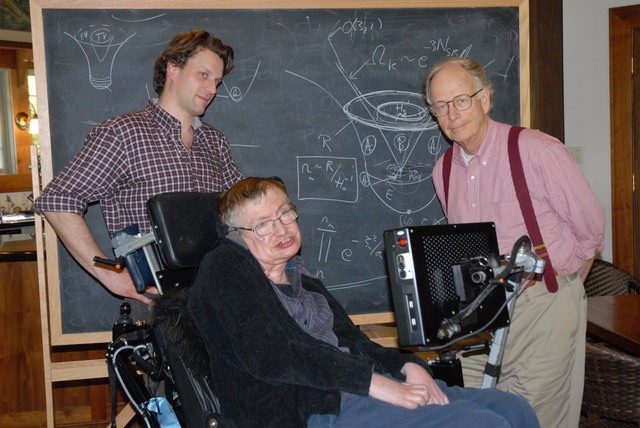}  
\caption{me Stephen and Thomas Hertog working together at John Mitchell's Cook's Branch Conservatory near Houston, Texas, (2009-11 approx.) :credit: Anna Zytkov. } 
\end{figure}

\begin{figure}h]
\includegraphics[width=5in]{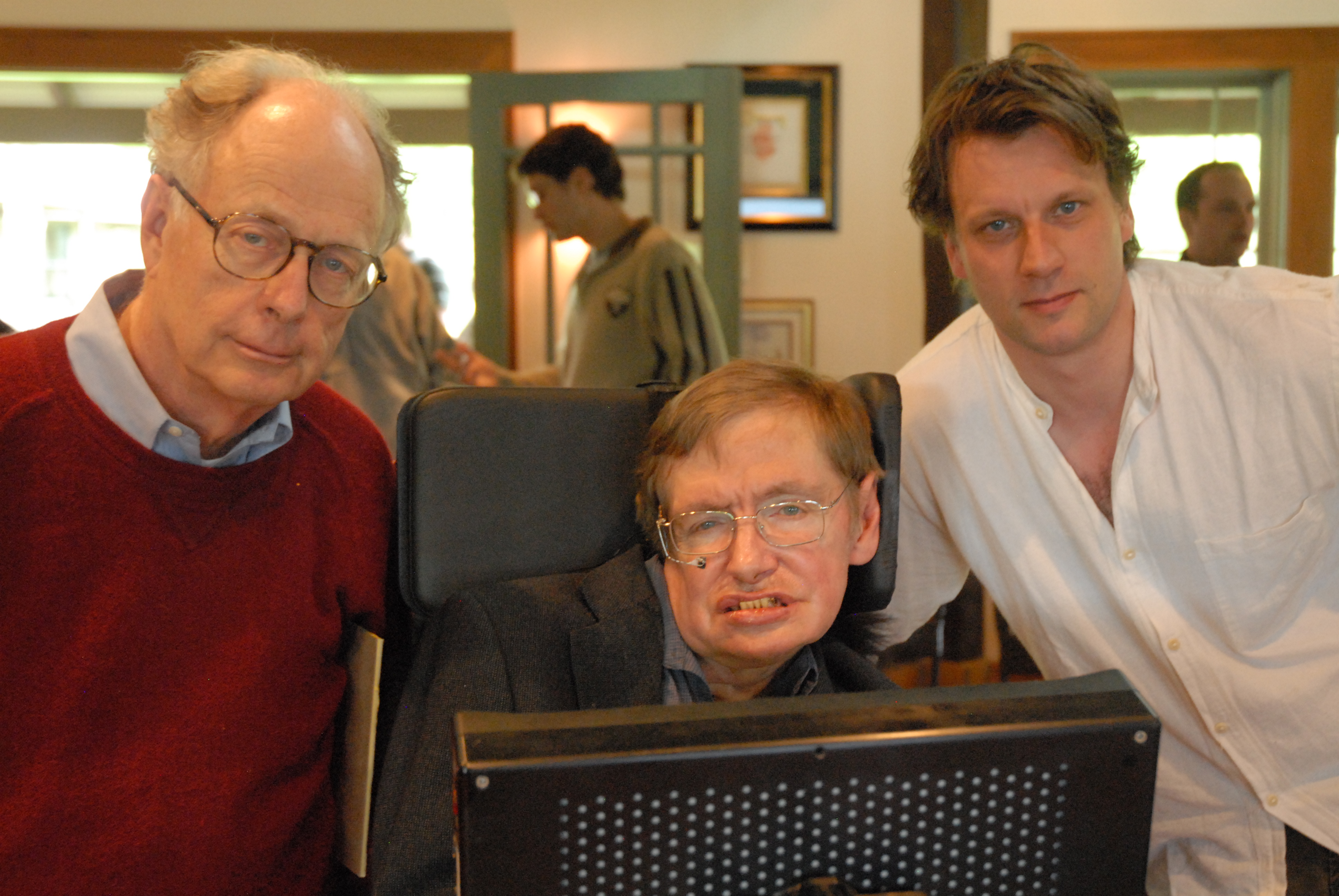}
\caption{me Stephen and Thomas Hertog working together at John Mitchell's Cook's Branch Conservatory near Houston, Texas, (2009-11 approx.) , credit Anna Zytkov}

\end{figure}

\begin{figure}[t]
\includegraphics[width=4in]{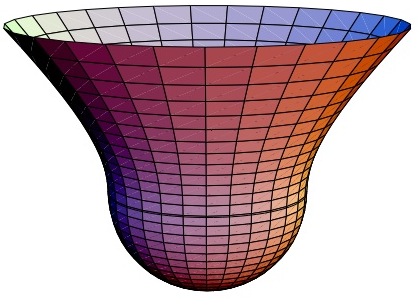}
\caption{The geometry of a two-dimensional slice through a saddle point contributing to the NBWF.  It has one boundary for arguments of the WAVU and no other boundary.} 
\end{figure}

\vspace{.2in}

\noindent{\bf Acknowledgments:}   Thanks to Graham Farmelo  for questions which led the author to think more clearly about the origins of the NBWF and write this essay. Thanks also to Thomas Hertog who read through it, and to the US National Science Foundation which  supported it under grant PHY-18-8018105.

\section{The Bibliography}

 \end{document}